\documentclass[epj]{svjour}
\usepackage{graphics}
\usepackage{epsfig}
\usepackage{amsmath}
\usepackage{xcolor}
\usepackage{mhchem}

\begin{document}
\title{Precision measurements of differential cross sections and analyzing powers in elastic deuteron-deuteron scattering at 65 MeV/nucleon}
\subtitle{}

\author{R.~Ramazani-Sharifabadi\inst{1,2}\thanks{reza\_ramazani@ut.ac.ir}\and 
        A.~Ramazani-Moghaddam-Arani\inst{3}\thanks{ramezamo@kashanu.ac.ir}\and
        H.R.~Amir-Ahmadi\inst{2}\and
        C. D.~Bailey\inst{4}\thanks{Present address: American Physical Society, 1 Physics Ellipse, College Park, MD 20240 USA}\and
        A.~Deltuva\inst{5}\and
        M.~Eslami-Kalantari\inst{6}\and
        N.~Kalantar-Nayestanaki\inst{2}\and
        St.~Kistryn\inst{7}\and
        A.~Kozela\inst{8}\and
        M.~Mahjour-Shafiei\inst{1}\and 
        H.~Mardanpour\inst{2}\and 
        J.G.~Messchendorp \inst{2}\and
        M.~Mohammadi-Dadkan\inst{2,9}\and
        E.~Stephan\inst{10}\and
        E. J.~Stephenson\inst{4}\and
        H.~Tavakoli-Zaniani\inst{2,6}
}                     

\institute{Department of Physics, University of Tehran, Tehran, Iran\and
KVI-CART, University of Groningen, Groningen, The Netherlands\and
Department of Nuclear Physics, Faculty of Physics, University of Kashan, Kashan, Iran\and
Center for Exploration of Energy and Matter, Indiana University, Bloomington, IN 47408 USA\and
Institute of Theoretical Physics and Astronomy, Vilnius University, Vilnius, Lithuania\and 
Department of Physics, School of Science, Yazd University, Yazd, Iran\and
Institute of Physics, Jagiellonian University, Krak\'ow, Poland\and 
Institute of Nuclear Physics, PAS, Krak\'ow, Poland\and 
Department of Physics, University of Sistan and Baluchestan, Zahedan, Iran\and 
Institute of Physics, University of Silesia, Chorz\'ow, Poland
}
\date{Received: date / Revised version: date}
%
\abstract{
  We present measurements of differential cross sections and analyzing powers
  for the elastic $^{2}{\rm H}(\vec d,d){d}$ scattering process. The data were obtained using a 130~MeV polarized deuteron beam. Cross sections and spin observables of the elastic scattering process were measured at the AGOR facility at KVI using two independent setups, namely BINA and BBS. The data harvest at setups are in excellent agreement with each other and allowed us to carry out a thorough systematic analysis to provide the most accurate   data in elastic deuteron-deuteron scattering at intermediate energies. The results
  can be used to confront upcoming state-of-the-art calculations in the four-nucleon
  scattering domain, and will, thereby, provide further insights in the dynamics
  of three- and four-nucleon forces in few-nucleon systems.
} 
\maketitle
\section{Introduction}
\label{intro}
\sloppy
Understanding the degrees of freedom that describe nuclear forces is of great importance
to make progress in nuclear physics. The first major breakthrough came in 1935 when Yukawa presented
the description of the nucleon-nucleon force by the exchange of massive mesons~{\cite{Yukaw}}
in analogy to the exchange of massless photons describing successfully the electromagnetic interaction.
More recently, various phenomenological nucleon-nucleon (NN) potentials have been derived based
on Yukawa's idea. Some of these potentials were successfully linked to the underlying
fundamental theory of quantum chromodynamics~\cite{coon,D1}. Precision measurements obtained from
nucleon-nucleon scattering data are strikingly well described by these modern NN potentials~\cite{MCM}.\\
\sloppy
\indent It is compelling to apply the high-precision NN potentials to systems composed
of at least three nucleons. Rigorous Faddeev calculations of the binding energy of the
simplest three-nucleon system, triton, underestimate the experimental data~{\cite{wir}}.
This observation shows that calculations based solely
on NN potentials are not sufficient to describe systems that involve more than two nucleons.
This has led to the notion of the three-nucleon force (3NF), a concept that
was introduced already in the early days of nuclear physics by Primakoff and Holstein~{\cite{prima}}.
Green's function Monte Carlo calculations based on the AV18 NN potential complemented with the
IL7 three-nucleon potential demonstrated the necessity of the 3NF to describe
the experimental data for the binding energies of light nuclei~{\cite{piep}}.
Moreover, rigorous Faddeev calculations based on modern NN potentials show large discrepancies with
cross section data in elastic nucleon-deuteron scattering. The inclusion of 3NF effects
partly resolves these deficiencies~{\cite{wita}}.
There are, by now, a large number of evidences revealing the importance of 3NF effects.\\
\sloppy
\indent In the last decades,
many nucleon-deuteron elastic~{\cite{kars01,kars03,kars05,sakai00,kimiko02,kimiko05,postma,hamid,kurodo,mermod,Igo,ald,Hos,Ela07,shimi,hatan,IUCF,Ahmad1,Ahmad2}}
and breakup~{\cite{st1,st2,st3,nasser,hos2,steph,myp1,myp2,myp3,tavakoli}} scattering experiments at various energies below the
pion-production threshold have provided an extensive database for the study 3NF effects.
The addition of 3NF effects, in particular the role of the $\Delta$ resonance,
reduces significantly the discrepancies between differential cross-section data and corresponding calculations excluding 3NF effects. The situation for spin observables is vastly different.
For instance, the inclusion of 3NF effects for the vector analyzing power of the elastic channel
at the intermediate energies gives a better agreement between data and theory, while for the tensor
analyzing power, $Re(T_{22})$, the discrepancies are not removed by adding 3NF effects in the model
~{\cite{nasser1}}. The inclusion of 3NF effects even deteriorates the agreement between model predictions and the data for the vector analyzing power of the proton in the proton-deuteron breakup
reaction at configurations that correspond to small relative energies between the two outgoing protons~{\cite{nasser1}}.
These observations imply that spin-dependent parts of 3NF effects are not yet well understood~{\cite{shi,nasser1}}.\\
\sloppy
\indent Although the three-nucleon (3N) system is the cleanest system to study 3NF effects since
only NN and 3N forces can contribute and observables can be calculated in an {\it ab-initio} manner,
the influence of 3NF effects are in general small in a 3N system.
Only at specific parts of the phase space in three-nucleon scattering processes, 3NF effects
become significant. A well-known example of such a phase space appears at scattering angles
corresponding to the minimum of the differential cross section in elastic N$d$ scattering~{\cite{wita,nemoto}}. In spin observables, a significant 3NF effect can also be seen for $pd$ break-up configurations corresponding to small relative energies between the two outgoing protons~{\cite{nasser1}}. Alternatively, and this is the focus of this paper, one may investigate the four-nucleon (4N) system
in which 3NF effects could be significantly enhanced~{\cite{nasser1}}.
Deuteron-deuteron scattering, as a 4N system, is a rich laboratory to study 3NF effects because of its
variety of final states, observables, and kinematical configurations.
Compared to the amount of available data in the 3N scattering domain, the database in the 4N system
is very limited. Most of the 4N data cover the very low-energy regime, below the three- and
four-body breakup threshold~{\cite{phill,vivi,fish}}. Although, calculations at these low energies are very
reliable, the effect of the 3NF is very small. Therefore, the low-energy realm is not the most attractive regime
to study rigorously the dynamics of 3NFs.\\
\sloppy
\indent  {\it Ab-initio} theoretical calculations for four nucleon systems are still limited to beam energies below 40 MeV~\cite{De1,De2,De3,De4,De5,De6,De7,De8}. At intermediate energies, below the pion-production threshold, the 4N experimental database is very scarce~{\cite{bech,Aldr,Garc}}.
Despite the fact that {\it ab-initio} calculations are still in development in this energy regime, the prospects
of studying the structure of 3N forces, and possibly higher-order four-nucleon force effects,
look promising~{\cite{Els,uzu}}. Recent theoretical approximations for deuteron-deuteron scattering are able to reasonably predict the experimental results in the quasi-free (QF) regime~{\cite{Delt2,Delt3}}. However, one should consider the final-state interactions of spectator neutrons to identify the QF limit correctly~\cite{myp1}. Besides, charge symmetry breaking studies (CSB) in $d+d\rightarrow \ce{^4{\rm He}}+\pi^{0}$ reaction reveal the necessity of theoretical calculations in $dd$ elastic scattering process to provide an unambiguous formulation of the initial-state interaction. In this energy regime, a single-scattering approximation is used in which one nucleon scatters from the opposite deuteron before it recombines to reform the original deuteron~\cite{Ed1,Micher}.\\
\sloppy
\indent This paper presents measurements of the differential cross section and spin observables in the
$^{2}{\rm H}(\vec d,d){d}$ elastic scattering process for a deuteron-beam energy of 65~MeV/nucleon.
The data were obtained by making use of a vector- and tensor-polarized deuteron beam that was provided by the
AGOR facility at KVI in Groningen, the Netherlands. Two experimental equipments, located at two different
beam lines, were used to measure independently the various observables in $^{2}{\rm H}(\vec d,d){d}$
scattering, namely the Big-Bite Spectrometer (BBS) and the Big Instrument for Nuclear-Polarization Analysis
(BINA). These setups bring complementary features: one (BINA) covering large phase space, particularly in $\phi$, using a liquid deuterium target leading to less background. The other (BBS) possesses an excellent momentum resolution, but with moderate coverage, using a solid CD$_{2}$ target with more precise knowledge on the target thickness at the cost of a larger background. These two sets of measurements combined have provided a good experimental database that can be used as a benchmark for future {\it ab-initio} calculations. This paper addresses the analysis of these two independent datasets.
The results presented here are the most precise and accurate data of the $^{2}{\rm H}(\vec d,d){d}$ process at
intermediate energies.
\section{Experimental setups}
\label{Exp}
\indent This experiment was performed with two different setups, BINA and BBS. In the following, details of both setups relevant for the present paper will be presented. Detailed discriptions are presented in~\cite{myth,BBSref}, respectively.
\subsection{Common source and accelerator facilties}
\indent The two experiments were conducted using AGOR facility at KVI. The measurement on BINA took place the week after the BBS data taking. BINA has the ability to identify and measure all reaction channels of the deuteron-deuteron scattering process simultaneously, while BBS  measures the hadronic channels with particles emerging from the two-body final states. Vector- and tensor-polarized (unpolarized) deuteron beams were produced by the atomic Polarized Ion
Source (POLIS) \cite{Fri,Krem} with nominal polarization values between 60-80\% of the theoretical values and
accelerated by the AGOR cyclotron to energies of 65 MeV/nucleon. The polarization of the deuteron beam was monitored for different periods of the experiment and found to be stable within statistical uncertainties~\cite{Ahmad4}.
\subsection{BINA}
\indent Figure~\ref{bin} shows a sketch of BINA. The setup consists of two parts, a forward wall and a backward ball. The forward wall consists of
a multi-wire proportional chamber (MWPC) to determine the scattering angles of the particles, twelve-vertically mounted plastic
$\Delta$E-scintillators with a thickness of 2~mm, and ten horizontally mounted E-scintillators with a thickness of 12~cm. The E-scintillators are
placed in a cylindrical shape where the target is positioned on the axial symmetry of the cylinder. Although, the $\Delta$E-E hodoscope provides
the possibility to perform particle identification, the information from the $\Delta$E detector was not used in this experiment. In a visual inspection after the experiment, these scintillators were observed to be damaged. Therefore, the $\Delta$E-E detectors could not provide the PID information for all scattering angles. Photomultiplier tubes (PMTs) were mounted on both sides of each E-scintillator. Signals from these PMTs are used to extract the energy and time-of-flight (TOF) of each scattered particle. The TOF information is used to perform PID. The MWPC covers scattering angles between 10$^\circ$ and 32$^\circ$ with a full azimuthal angle coverage and up to 37$^\circ$
with a limited azimuthal angle coverage. The MWPC has a resolution of $0.4^{\circ}$ for the polar angle and between $0.6^{\circ}$ and $2.0^{\circ}$ for the
azimuthal angle depending on the scattering angle. The detection efficiency of the MWPC for deuteron with energies corresponding to the reaction of interest is typically $98\pm1\%$~{\cite{Ahmad3}}. The
backward ball of BINA is made of 149 phoswich scintillators that were simultaneously used as detector and scattering chamber with a scattering-angle coverage
between $40^\circ$ and $165^\circ$ and nearly full azimuthal coverage. For more details on BINA, we refer to \cite{myth,hos3}.\\
\indent The deuteron beam, with a typical current of 4 pA, bombarded a liquid-deuterium target that was mounted inside the scattering chamber of
BINA~{\cite{nasser3}. The thickness of the target cell was 3.85~mm with an uncertainty of 5\%. The scattering angles, energies, and (partly) time of flights of the
final-state deuterons were measured by the multi-wire proportional chamber (MWPC) and scintillators of BINA.
A Faraday cup was mounted at the end of the beam line to monitor the beam current throughout the experiment. The current meter of the Faraday cup
was calibrated using a current source with an uncertainty of 2\%~{\cite{Ahmad4}}. A small offset in the readout of the current was observed with a value around 0.28 $\pm$ 0.13~pA, see Sec.~\ref{Exres}.
\vspace{-1cm}\begin{figure}[t]
 \centering 
\epsfxsize=8.5 cm
\epsfbox{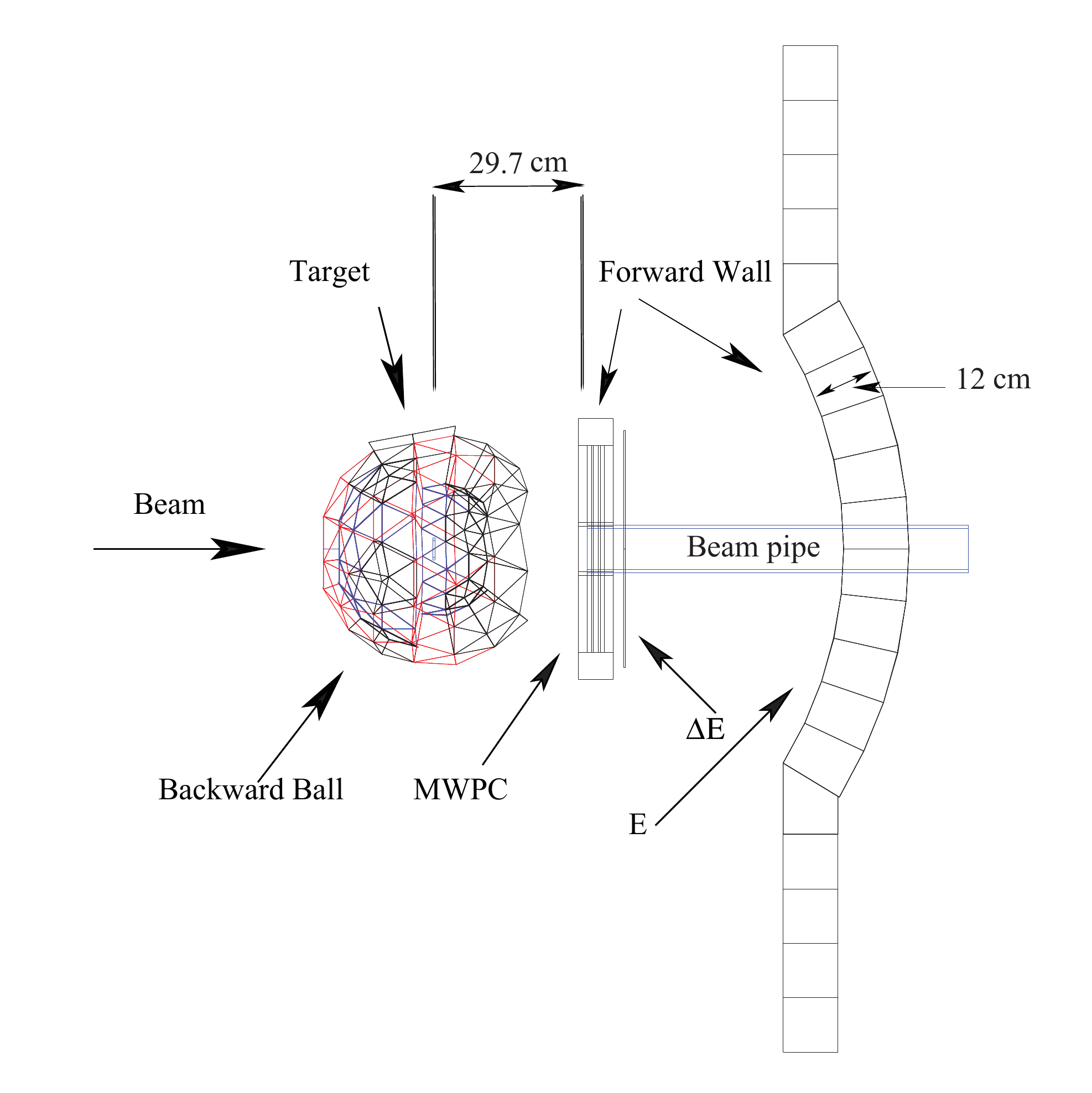} 
\vspace*{0.0cm}
\caption{A sketch of the various components of the BINA setup. The elements on the right side show a side view of the forward part of BINA,
  including the multi-wire proportional chamber (MWPC), an array of twelve thin plastic $\Delta$E-scintillators followed by ten thick segmented
  E-scintillators mounted in a cylindrical shape. On the left side, the backward part of BINA is depicted composed of 149 phoswich scintillators
  glued together to form the scattering chamber.}
\label{bin}
\end{figure}\\
\subsection{BBS}
\indent The Big-Bite Spectrometer (BBS) is a QQD-type magnetic spectrometer with a $K$-value of 430 MeV and a solid angle of up to 13 msr. By changing the position of the quadrupole doublet with respect to the dipole magnet, while the distance between the object (target) and the dipole remains the same, the momentum-bite acceptance can be changed from 13 to 25$\%$, the solid angle changes from 13 to 7 msr, simultaneously. The BBS consists of a scattering chamber containing a target ladder, a large slit wheel containing several entrance apertures (including a sieve slit for angle reconstruction), two sets of quadrupole magnets for beam focusing, a large dipole magnet for momentum selection, two sets of x-u plane wire-chamber detectors, and a scintillator plane which is used to generate the event trigger. A diagram of the BBS is shown in Fig.~\ref{bbsfig}.\\
\indent In the BBS setup, different thick or thin sets of CD$_{2}$ and carbon targets were used for different ranges of lab angles. The carbon targets were used in the forward range of spectrometer angles to be able to subtract the background generated by deuterons elastically scattered from carbon in the CD$_2$ target.
For large angles ($\geq$ 15$^\circ$), several layers of solid CD$_{2}$ were combined, resulting in a total thickness of 45.15 $\pm$ 2.26 mg/cm$^{2}$. For small angles (4$^\circ$ and 6$^\circ$) the CD$_2$ target thickness was 10.49 $\pm$ 0.52 mg/cm$^2$. The thickness of the carbon target for large angles ($\geq$ 15$^\circ$) was 46.80 $\pm$ 0.65 mg/cm$^2$; for small angles it was 14.2 $\pm$ 0.20 mg/cm$^2$. The scattering chamber of the BBS consisted of a large cylindrical chamber containing the targets and essentially forming the pivot point around which the device covers the scattering angles between 4$^\circ$ and 48$^\circ$ during the data taking. For the beam integration, a large copper Faraday cup was used for the angles larger than 15$^\circ$, where the unscattered beam could hit the wall of the scattering chamber. For small scattering angles (less than 15$^\circ$), the unscattered beam was within the acceptance of BBS and entered the region of the quadrupole magnets. Therefore, a separate Faraday cup is mounted between the quadrupole magnets Q1 and Q2 for the beam integration in this region. A detailed description of the BBS setup is presented in~\cite{NIMBBS}.\\
\vspace{-1cm}\begin{figure}[t]
 \centering 
\epsfxsize=7.5 cm
\epsfbox{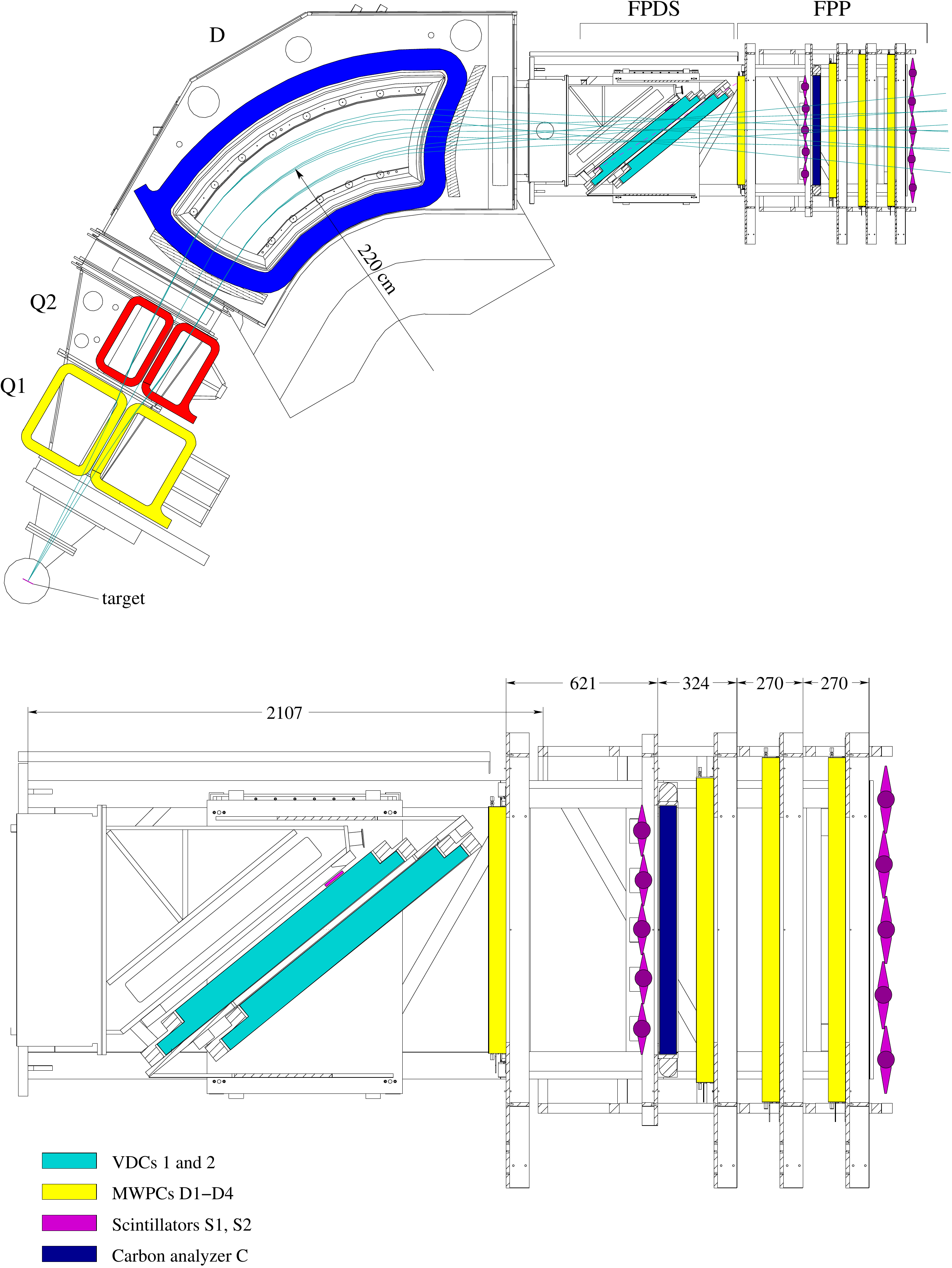} 
\vspace*{0.0cm}
\caption{A sketch of the main features of the BBS setup.}
\label{bbsfig}
\end{figure}
\section{Event selection and data analysis method}
\label{Ana}
\indent In this section, the analysis procedures of both experiments related to the BINA and BBS setups are described separately. Detailed discriptions are presented in~\cite{myth,BBSref}, respectively.
\subsection{BINA}
\indent During data taking with BINA, various hardware triggers with different down-scale factors were implemented that were dedicated to a specific hadronic
final state in deuteron-deuteron scattering. To select events originating from elastically scattered deuterons, two triggers were of importance.
The first one, referred to as the coincidence trigger, registered events for which there was at least one signal from the forward wall scintillators in coincidence with at least
one signal originating from the backward wall. This trigger was down-scaled by a factor two. The hardware thresholds for detection of a particle were typically set around 1~MeV. Although, with the coincidence trigger we were able to cover a large part of the angular distribution of the $^{2}{\rm H}(\vec d,d){d}$ reaction, whereby both deuterons in the final state were detected, we observed a significant drop in the detection efficiency for low-energetic deuterons that scatter towards the backward ball due to energy losses of those particles in the liquid-deuterium target. The data selected with the coincidence trigger were used to extract the spin observables, since detection inefficiencies cancel out in the analysis.
To extract the differential cross section, we exploited a second trigger the so-called, ``single trigger''. This trigger, down-scaled by a factor
256, was built from a logical OR of all the discriminated signals of the scintillators of BINA, and, thereby, not biased on the response of the
backward ball.

\indent The data from the coincidence trigger were calibrated and further preprocessed by requiring that the relative angles of the reconstructed
particles hitting the forward wall and backward ball match the correlation that is expected from kinematical considerations for the elastic
deuteron-deuteron scattering process. Cuts were applied to meet a relative opening angle of $83^\circ$ and a coplanar configuration with respect to
the azimuthal angles. After applying these angular cuts with a window of $\pm 20^\circ$, a major reduction (around $75\%$) of backgrounds from other hadronic final
states, such as breakup and nucleon-transfer reactions, was obtained. Figure~\ref{corre} shows the correlation between energy and scattering angle
of deuterons detected in the forward wall of BINA after the aforementioned event selection. The solid line represents the expected kinematical
locus for the elastic deuteron-deuteron scattering. As seen, elastically scattered deuterons can easily be observed and distinguished from background
channels. The data below the elastic events reveal another clear correlation which has been identified as events belonging to the neutron-transfer
channel, $^{2}{\rm H}(\vec d,\ce{^3{\rm H}}){p}$.\\ 
\hspace{-1cm}\begin{figure}[t]
 \centering 
\epsfxsize=8.8cm
\epsfbox{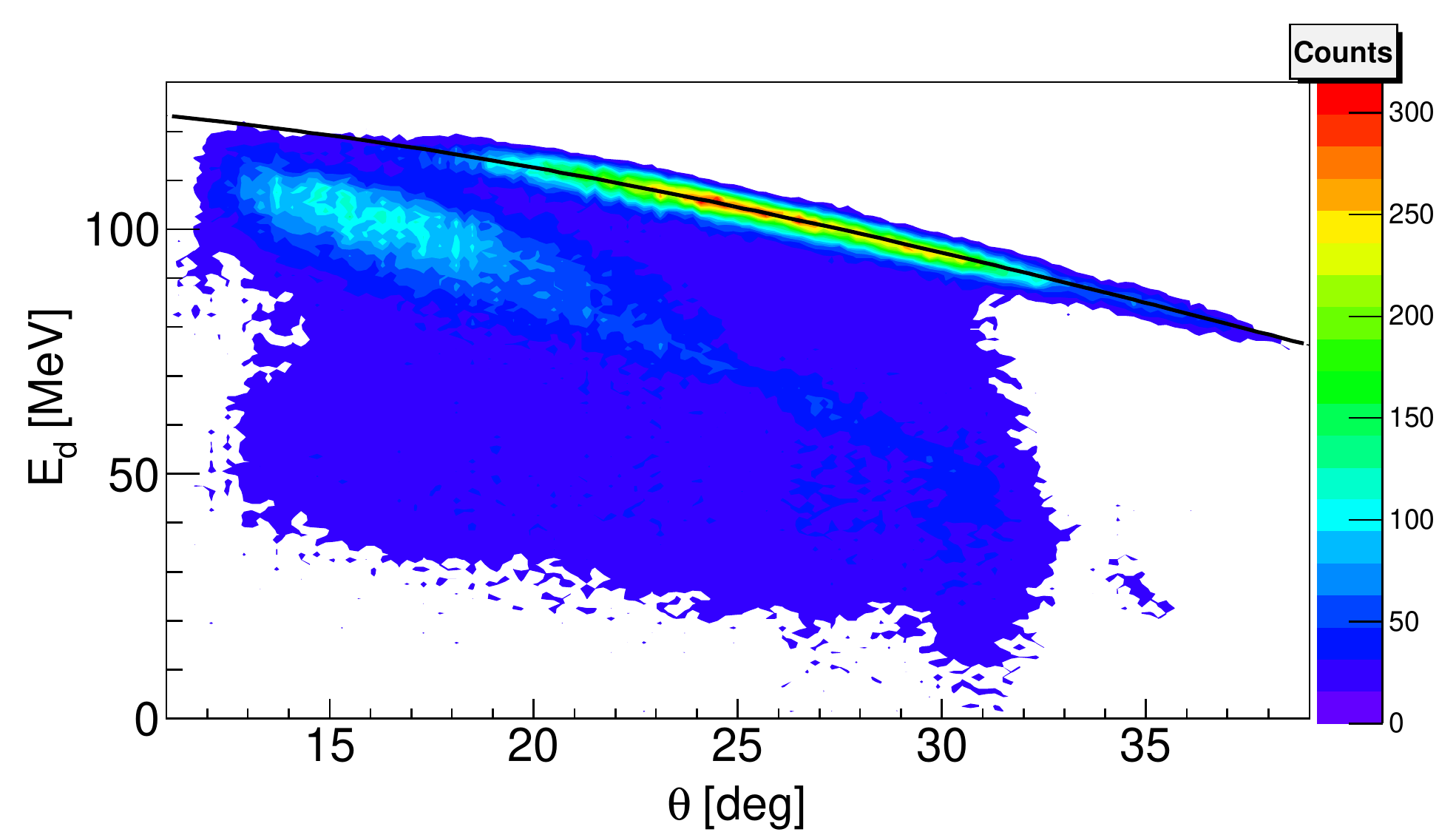} 
\vspace*{-0.5cm}
\caption{
  The correlation between the reconstructed energy and scattering angle of the particles that were detected in the forward part of BINA with a
  coincidence requirement with the backward ball. The solid line represents the kinematical locus for the elastic deuteron-deuteron scattering
  process.}\label{corre}
\end{figure}
\indent To count the number of events that originate from the elastic process, the center-of-mass energy for each reconstructed particle is
calculated from its energy deposit and scattering angle, and a corresponding histogram is generated in intervals of 2$^\circ$ of the scattering
angle and separated for the various polarization states of the beam. Figure~\ref{proj} depicts the center-of-mass energy distribution that has been
obtained using the single trigger. The upper spectrum shows the raw response after calibration and for particles that scatter to $26\pm1^\circ$ but
without any further conditions. For the lower spectrum, a coincidence with the backward ball was required in addition using the kinematical cuts
discussed earlier but from data taken with the single trigger. The solid lines are the result of a fit through the data based on a
Gaussian-distributed signal combined with a 5$^{th}$-order polynomial representing the backgrounds. The background component of the fit is indicated by
the dashed lines. A clear peak can be observed in both cases, corresponding unambiguously to the channel of interest. 
The difference between the integrals of the signal distributions before and after applying the coincidence condition excluding inefficiencies of the ball is less than 2$\%$. The coincidence requirement
reduces significantly the background contribution. Monte Carlo simulations showed that the remaining background is mostly due to hadronic
interactions of elastically scattered deuterons in the scintillator.
\hspace{-1cm}\begin{figure}[t]
 \centering 
\epsfxsize=9.2cm
\epsfbox{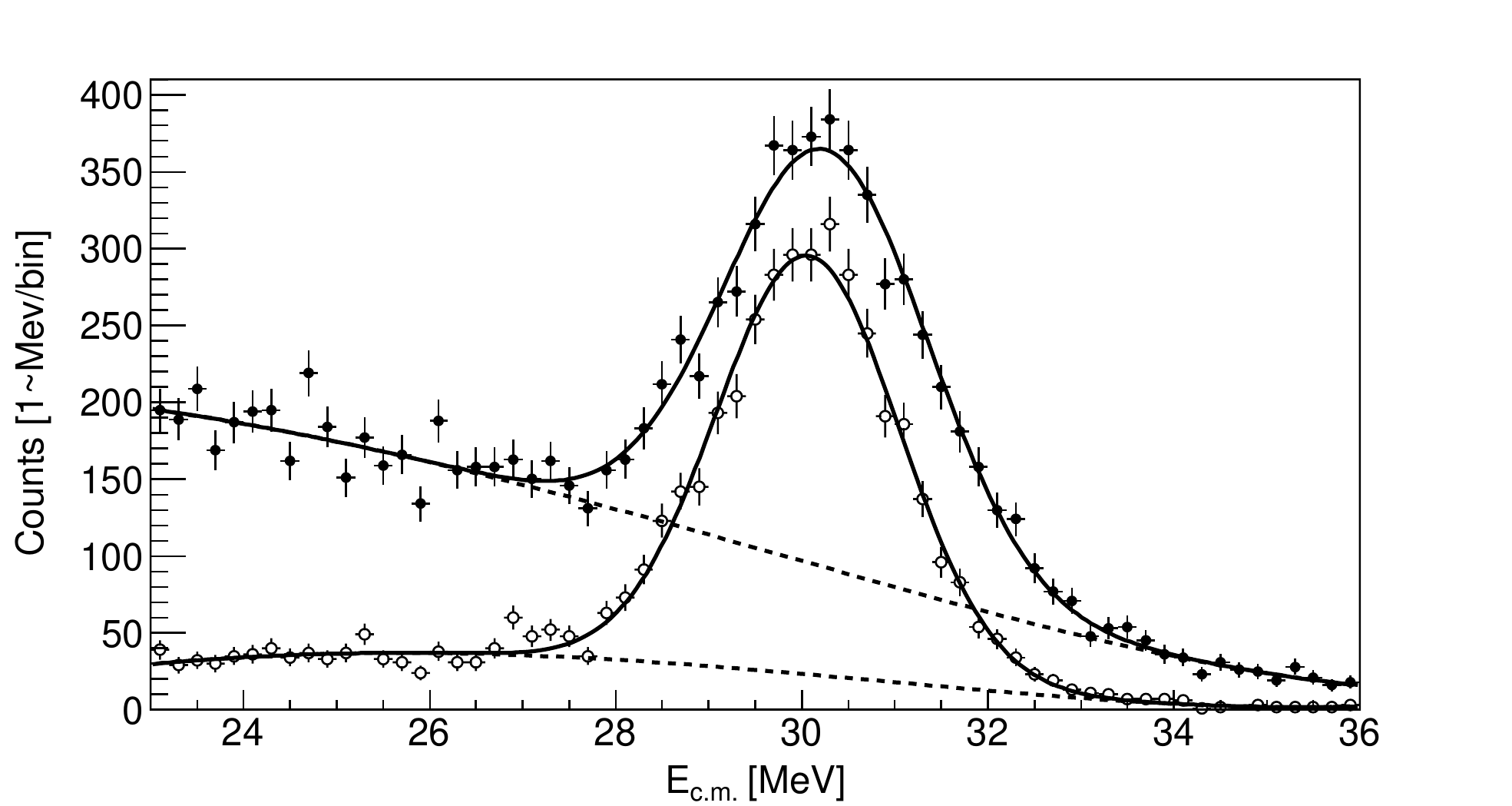} 
\vspace*{-0.5cm}
\caption{Spectrum of the center-of-mass energy of particles hitting the E detectors of the forward wall. Data are obtained using the single trigger.
  The scattered particles are confined to polar angles of $26\pm1^\circ$. For the lower spectrum, a coincidence condition is imposed in the event
  selection. The solid lines show the results of a least-$\chi^2$ fit based on a Gaussian (signal) and a 5$^{th}$-order polynomial (background)
  distribution. The background contribution is indicated by the dashed lines. The $\chi^2$/64 of the fit is 1.4 for the upper spectrum and 1.2
  for the lower one.}\label{proj}
\end{figure}

\indent  To extract cross sections, the number of counts passing the kinematical criteria has been corrected for efficiencies of the system such as
live-time, MWPC efficiencies, hadronic interactions, and the down-scale factor that comes from triggers. The average live-time of the data
acquisition of BINA is around $40\%$.\\
\indent Events from the elastic reaction that suffered from hadronic interactions do not give a clear peaking structure in the energy spectrum, and are, therefore, not easily separated from other background channels, we did not count these events and corrected for their loss. Using a GEANT3-based Monte Carlo simulation, we estimated a loss of $16\pm2\%$ for the energy range of interest. The cross sections are corrected for this effect accordingly.\\
\indent Vector and tensor polarized beams make it possible to measure spin observables. Using parity conservation, the cross section for
$^{2}{\rm H}(\vec d,d)d$ reaction is given by the following equation~{\cite{ohl}}:
\begin{eqnarray*}
\frac{\sigma(\theta,\phi)}{\sigma_{0}(\theta)}=k\Bigg[1+\frac{3}{2} p_{Z}A_{y}(\theta)\cos(\phi)-\frac{1}{4}p_{ZZ}A_{zz}(\theta) \\
+\frac{1}{4} p_{ZZ}\Big(A_{zz}(\theta)+2A_{yy}(\theta)\Big)\cos(2 \phi)\Bigg],~~(1)\label{eq1}
\end{eqnarray*}
where $\theta$ and $\phi$ are polar and azimuthal angles of the scattered deuteron, respectively. $A_{y}$ is the vector analyzing power, while $A_{zz}$ 
and $A_{yy}$ are the tensor analyzing powers. $p_{Z}$ ($p_{ZZ}$) represents the vector (tensor) polarization of the beam.
$\sigma$ ($\sigma_{0}$) is the effective cross section obtained for data taken with (un)polarized beam. These effective cross sections correspond to the number of counts normalized by the accumulated and dead-time corrected charge. Please note that in first order, the efficiencies cancel by taking the ratio between $\sigma(\theta,\phi)$ and $\sigma_{0}(\theta)$. Finally, $k$ is a normalization factor and should be equal to one in the ideal case.
Considering $k$ as a free parameter, it fluctuates around one with a value of $k=1.00\pm0.03$ that is considered as a systematic uncertainty for the normalization procedure.
 Experimentally, however, we evaluated possible systematical differences in the extraction of the effective cross sections $\sigma(\theta,\phi)$ and $\sigma_{0}(\theta)$ accommodated in $k$. These may be due to small differences in detection efficiencies or beam-current measurements between data taken with unpolarized and polarized beams. For the extraction of the analyzing powers, we analyzed data taken with the coincidence trigger and enforcing the selection criteria as described above. We note that the background using the coincidence conditions is very small.\\
\indent We extracted the analyzing powers with two different methods which both lead to compatible results within the uncertainties. In the first method; we assume that the beam polarization is purely vector ($p_{Z} \neq 0$ and $p_{ZZ} = 0$) or purely tensor ($p_{Z} = 0$ and $p_{ZZ} \neq 0$). Therefore, in Eq.~\ref{eq1}, the corresponding terms are kept and the other terms are set to zero.
 As can be seen in~Eq.~\ref{eq1}, the asymmetry ratio of polarized to un-polarized cross section is a function of $\cos\phi$ ($\cos2\phi$) for the case of pure vector (tensor) polarized beam, see Fig.~\ref{Asymm}. Therefore, vector analyzing power, $A_{y}$, is extracted from the amplitude of $\cos\phi$. In the same way, tensor analyzing powers of $A_{zz}$ and $A_{yy}$ are extracted from the off-set of $\cos2\phi$ from one and its amplitude, respectively. To estimate the systematic uncertainty due to the possible impurity in the vector- (tensor-)polarized beam, the second method is applied. 
 In the second method, we suppose that the pure-vector (tensor) polarized beam is not actually a pure-vector (tensor), ($p_{Z} \neq 0$ and $p_{ZZ} \neq 0$). In other words, the pure-vector (tensor) polarized beam is contaminated with another polarization, say the tensor (vector) polarization. Therefore, Eq.~\ref{eq1} including all the terms is used to fit to the asymmetry ratio of polarized to un-polarized cross section beam for vector and tensor analyzing powers. As described before, the analyzing powers can again be extracted from the amplitudes of the $\cos\phi$ and $\cos2\phi$ functions as well as the offset of $\cos2\phi$ function from one. The difference between the two results is considered as the systematic uncertainty due to the possible impurity in the beam polarization. The results of the first method is considered as the final results, see Sec.~\ref{Exres}.\\
\hspace{-1cm}\begin{figure}[t]
 \centering 
\epsfxsize=9.8cm
\epsfbox{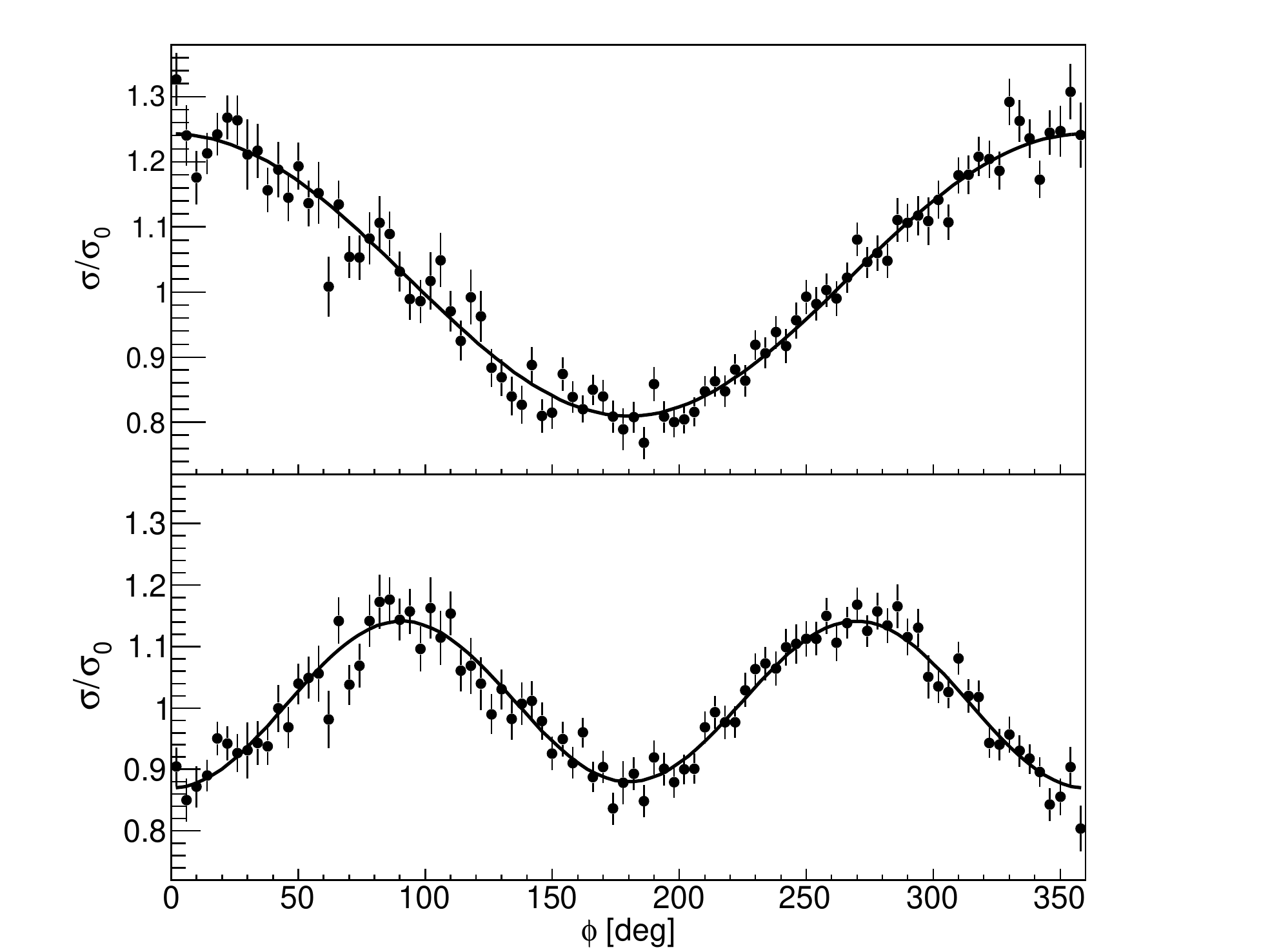} 
\vspace*{-0.5cm}
 \caption{Asymmetry ratio of cross section for polarized over un-polarized beam as a function of $\phi$ for a pure-vector polarized beam (top panel) and pure tensor polarized beam (bottom panel. Scattering angle of elastically scattered deuteron is $26\pm1^\circ$. The reduced $\chi^2$ for the top (bottom) panel is 1.04 (0.97).}\label{Asymm}
\end{figure}
\indent To verify the procedure of extracting the differential cross sections and analyzing powers of the $^{2}{\rm H}(\vec d,d)d$ reaction, we measured and analyzed the ${\rm H}(\vec d,dp)$ reaction as well. The same procedure was used to analyze the data of the well-studied ${\rm H}(\vec d,dp)$ reaction which were obtained using a CH$_{2}$ target and with the same setup and beam conditions as was applied in the study of the $^{2}{\rm H}(\vec d,d)d$ reaction.\\
%
%

\indent Differential cross sections and analyzing powers for the reaction ${\rm H}(\vec d,dp)$ are presented in Fig.~\ref{DP}. In each panel, the
results of this analysis are represented by filled circles. The error bars indicate the statistical uncertainties and the gray bands represent the systematical errors. A detailed description of the related systematic uncertainties is presented in~\cite{Ahmad1}. The open triangles show the results of cross sections measured at RCNP~\cite{shimi}.
The open circles and filled triangles show the analyzing powers data taken at KVI, \cite{Ahmad1,Ela07}, and open rectangles are those taken at
RIKEN \cite{Hos}. The solid curves show the results of a coupled-channel calculation by the Hannover-Lisbon theory group based on the CD-Bonn
potential including the Coulomb interaction and an intermediate $\Delta$-isobar \cite{Delt1}. The dotted lines represent results of a similar calculations
by excluding the $\Delta$-isobar. We note that the 3NF effects are predicted to be small and, therefore, the results of the presented Faddeev calculations based on the high-precision NN potential are expected to accurately describe the experimental data. In addition, the results are compared with the results of a rough approximation based on the lowest-order terms in the Born series expansion of the Alt-Grassberger-Sandhas (AGS) equation for a three-nucleons interaction using CD-Bonn+$\Delta$ potential (the dashed lines). The comparison shows that the Born approximation is not very good in three-body systems at this energy, and therefore, we do not expect that such an approach will provide a good description in the four-body scattering process; see Sec.~\ref{Exres}. It is worth noting that the quality of the Born approximation improves with increasing the energy and/or at small scattering angles as the lowest-order terms become dominant in all observables. Our measurements for the ${\rm H}(\vec d,dp)$ reaction are in excellent agreement with previously published data and with state-of-the-art calculations, lending, thereby, confidence in the analysis procedure and our estimates of systematic uncertainties.
\begin{figure*}[!t]
\centering
\resizebox{13cm}{!}{\includegraphics[angle = 0,width =1\textwidth]{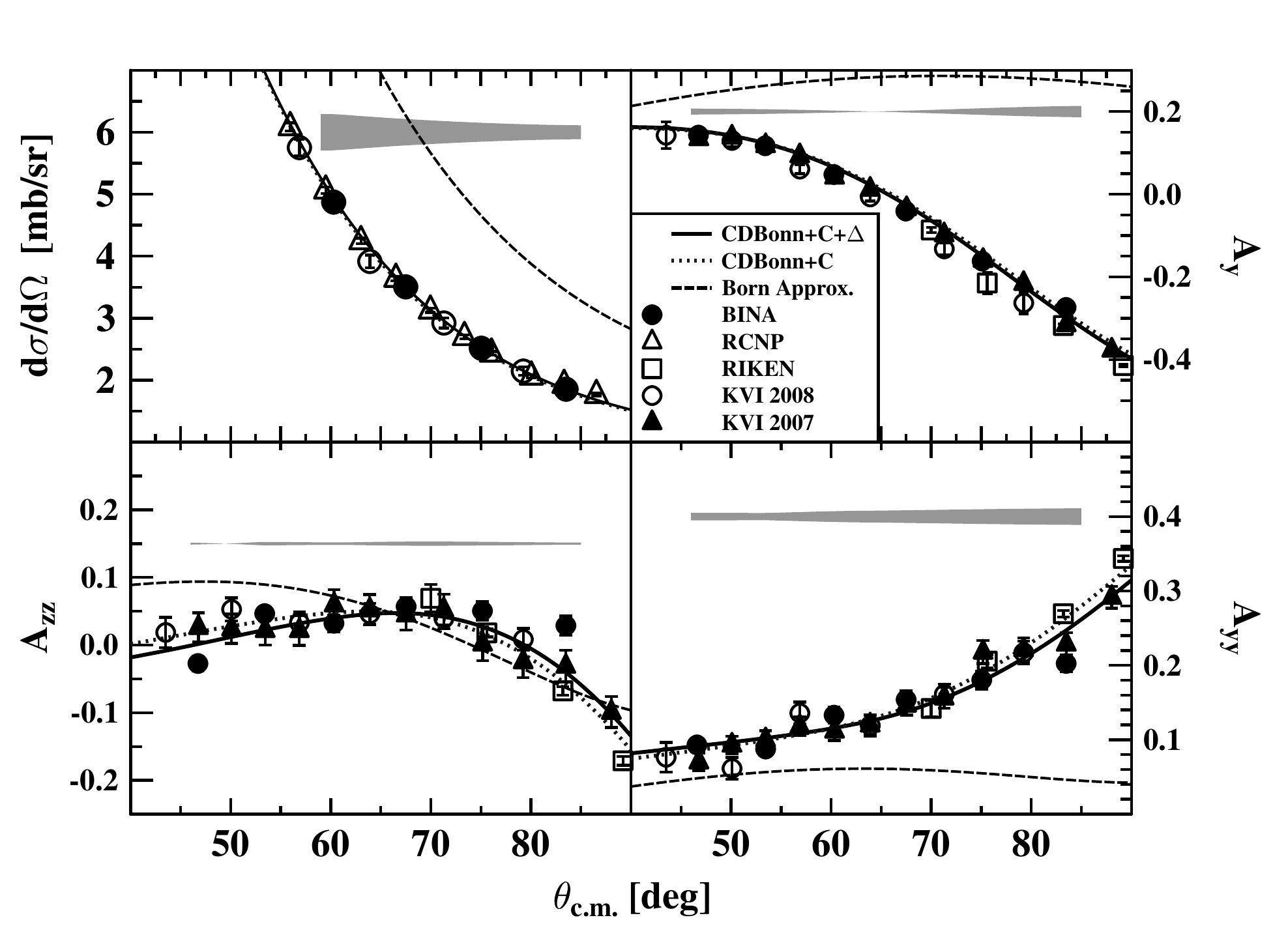}}
\vspace*{-0.2cm}
\caption{Differential cross section and analyzing powers of the elastic channel of the reaction ${\rm H}(\vec d,dp)$ that were taken with a deuteron beam of 65~MeV/nucleon. In each panel, the data taken with BINA are indicated with filled circles whereby the error bars are statistical. The open triangles show the cross section results obtained at RCNP~\cite{shimi}. The open circles and filled triangles show the analyzing powers data taken at KVI~\cite{Ahmad1,Ela07}, and open rectangles are those obtained at RIKEN \cite{Hos}. The solid curves show the results of a coupled-channel calculation by the Hannover-Lisbon theory group based on the CD-Bonn potential including the Coulomb interaction and an intermediate $\Delta$-isobar~\cite{Delt1}. The dotted lines represent results of a similar
calculation by excluding the $\Delta$-isobar. The dashed lines represent the results of a rough approximation based on the lowest-order terms in the Born series expansion of the Alt-Grassberger-Sandhas (AGS) equation using CD-Bonn+$\Delta$ potential. The gray band shows the systematic error (2$\sigma$) in each panel,}
\label{DP}
\end{figure*}
\subsection{BBS}
\indent In the following, the analysis procedure of the BBS data is described. Details of the analysis methods related the BBS data are presented in~\cite{BBSref}. \\
\indent The differential cross section and spin observables were extracted at various scattering angles by counting elastically scattered deuterons for various polarization states of the beam. To access different scattering angles, the spectrometer was moved around the target. The quadrupole and dipole fields were changed according to the kinematics of the related reaction to focus and bend the particles of interest and bring them to the detector plane. In this case, one focal point was produced via a combination of quadrupole and dipole fields for a scattered particle with a given momentum. Therefore, the solid angle spanned by particles, as they scatter from the target inside the scattering chamber, were determined by a defining aperture in front of the spectrometer. For this purpose, a ``seive slit'', an aperture fitted into the slit wheel of the BBS containing several pre-drilled holes, was used during several runs of the experiment. With this slit system, the optical coefficients of BBS were fitted and the system was, therefore, calibrated for various settings.\\
\indent The main background sources are the events including deuterons elastically or inelastically scattered from Carbon. These events are appeared in the detector plane along with the events of interest.
To subtract the background, we applied two techniques. For the runs with no discernible background structure due to the Carbon in the CH$_{2}$ target, the procedure of background subtraction is similar to that described for the event selection in BINA. For the runs in which a clear background structure due to the Carbon in the CH$_{2}$ target was present, the separate Carbon data from the Carbon target which were taken during the experiment were used. For each of these runs the corresponding Carbon data (i.e. data which were taken with exactly the same spectrometer settings and beam energy, but with a solid carbon target) were analyzed using the same parameters as the reaction data of interest. Finally, to obtain the differential cross section for each of the five beam polarization states, the extracted number of counts after background subtraction is corrected for the efficiencies of the system such as live-time, and wire chamber efficiencies.\\
\indent By knowing the polarized and unpolarized cross sections for each of the five beam polarization values, we could then calculate the unpolarized cross section and analyzing powers $A_{y}$ and $A_{yy}$ using Eq.~\ref{eq1} through a simple matrix inversion. We have five equations and only three unknowns: the unpolarized cross section $\sigma_0, A_y,$ and $A_{yy}$. Therefore, the analyzing powers are obtained from the polarized cross sections using a matrix inversion, and their statistical errors determined using standard error propagation techniques. Generally, there were almost always five good polarized cross sections available, and therefore, this was an over-determined system; however for a few runs only three or four polarization states were available, in which case the matrix inversion was reduced to only include the existing polarized cross sections~\cite{BBSref}. 
\section{Systematic uncertainties}
\indent The common systematic uncertainties between the two experiments as well as those specifically for BINA and BBS setups are separately presented in the following subsections.
\subsection{Common sources of systematic errors}
The main common source of systematic error comes from the uncertainty in the $A_{y}$  measurements in the ${\rm H}(\vec d,dp)$ reaction to extract the polarization that is around 4.5$\%$. 
The results of $A_{yy}$  measurements obtained from BBS are also used to estimate an offset in the readout of the current. The offset was determined by minimizing the reduced $\chi^2$ whereby an offset in the current is introduced as a free parameter using the comparison between the results of the $A_{yy}$ from the elastic channel of $dd$ scattering coming from the BINA and those coming from the BBS~{\cite{BBSref}}. The error is obtained by evaluating the $\chi^2$ distribution as a function of the offset. The intersection point of this distribution with a $\chi^2$ value that is one unit larger than its minimum has been used to determine the uncertainty in the offset. The systematic error arising from the measurement of the beam current using a Faraday cup leads to a small offset of 0.28 $\pm$ 0.13~pA in the readout of the current. \\
\subsection{BINA}
\indent One of the systematic uncertainty in the cross section measurement is attributed to the thickness of the liquid deuterium target. We estimated a corresponding error of $5\%$ due to the bulging of the cell. The size of bulging was first estimated via a measurement of the target thickness as a function of pressure at room temperature. The actual target thickness was obtained by comparing the cross section measurements at KVI between solid and liquid targets and the difference is considered as the uncertainty due to the thickness measurement. Other systematic uncertainties come from the beam luminosity using a precision current source ($2\%$), the MWPC efficiency for deuterons which was obtained using an unbiased and nearly background-free data sample of the $pd$ elastic scattering process ($1\%$), and the errors in the correction factor for losses due to hadronic interactions in the detector. For deuterons, this error is extracted from the difference between the measured and simulated deposition of deuteron energy in the forward wall of BINA ($2\%$)~\cite{Ahmad4}. The uncertainty of the extraction of the differential cross sections due to the offset current is around 5$\%$. To estimate the systematic uncertainty due to the background model, we used the 3$^{th}$ and 7$^{th}$ orders of polynomial fit-functions instead of the 5$^{th}$ order polynomial representing the backgrounds. The maximum difference between the results are considered as the systematic uncertainty due to the background model which is around 4.5$\%$. \\
\indent The polarization of deuteron beam was monitored with a Lamb-Shift Polarimeter (LSP)~\cite{lsp} at the low-energy beam line and measured with BINA after beam acceleration at the high-energy beam line by measuring the asymmetry in the elastic deuteron-proton scattering process~\cite{ibp}. The vector and tensor polarizations of the deuteron beam of BINA were found to be $p_{Z}=-0.601\pm 0.029$ and $p_{ZZ}=-1.517\pm 0.032$, respectively, whereby the errors include uncertainties in the analyzing powers in elastic deuteron-proton scattering. It should be remarked that only negative polarization states were used, since the number of events obtained with that polarization state is much larger than those obtained with the opposite polarization state. The errors were extracted employing a constant-line fit through the measured polarization values as a function of center-of-mass angle.
 In the case of measuring analyzing powers, a systematic uncertainty comes from the normalization procedure by considering the $k$ factor in Eq.~\ref{eq1} as a free parameter. This error turned out to be around $3\%$. Moreover, the maximum shift in the results of $A_{y}$, $A_{yy}$, and $A_{zz}$ due to the offset current is around 0.01, 0.035, and 0.08, respectively, while the measured values of these observables vary between $-0.07$ to +0.35, $-0.04$ to +0.22, and $-0.06$ to +0.3, respectively. The systematic error due to the possible impurity in the beam polarization is negligible for the vector analyzing powers and estimated to be about 0.01 (absolute) for the tensor analyzing powers.\\
\subsection{BBS}
\indent Systematic errors in the measurement of the differential cross sections originate mainly from the errors in the knowledge of the target thickness and the calibration of the Faraday cup. As was already stated, the error in the target thickness for the elastic $dd$ reaction was around $5\%$.
The errors in the areal density measurements, which is the mass of the material divided by its area with the unit of $\rm {mg/cm^2}$, come from both mass measurement errors and those from the measurements of the size of the target. The error in the calibration of the Faraday cup was estimated to be $0.5\%$. These components, added in quadrature, were applied as an overall scale factor systematic for the cross sections (yielding a total normalization error of about $5\%$ for the elastic $dd$ reaction).\\
\indent The polarization states of the deuteron beam of the BBS  were measured with the Ion-Beam Polarimeter (IBP) and found to be as follows: vector plus ($0.538 \pm 0.029$), vector minus ($- 0.621 \pm 0.030$), tensor plus ($0.671 \pm 0.04$), and tensor minus ($- 1.633 \pm 0.035$). 
The main source of systematic uncertainty in this method comes from the analyzing powers measurements in the elastic $d + p$ reaction while using IBP. The polarization values for each state were measured at different beam energy ranges and found to be consistent within the statistical uncertainties~\cite{BBSref}. The main sources of systematic error for the analyzing powers include the uncertainty due to beam polarization measurements ($p_{y}$ and $p_{yy}$), and the total calibration error.
 The calibration errors for ($A_{y}, A_{yy}$) are found to be around ($1.2\%, 1.7\%$). These errors introduce an overall scale factor, since the beam polarization and initial polarimetry calibration apply to all angles. Details of systematic studies are presented in~\cite{BBSref}.
\section{Experimental results}
\label{Exres}

\begin{figure*}[ht]
\centering
\resizebox{14cm}{!}{\includegraphics[angle = 0,width =1\textwidth]{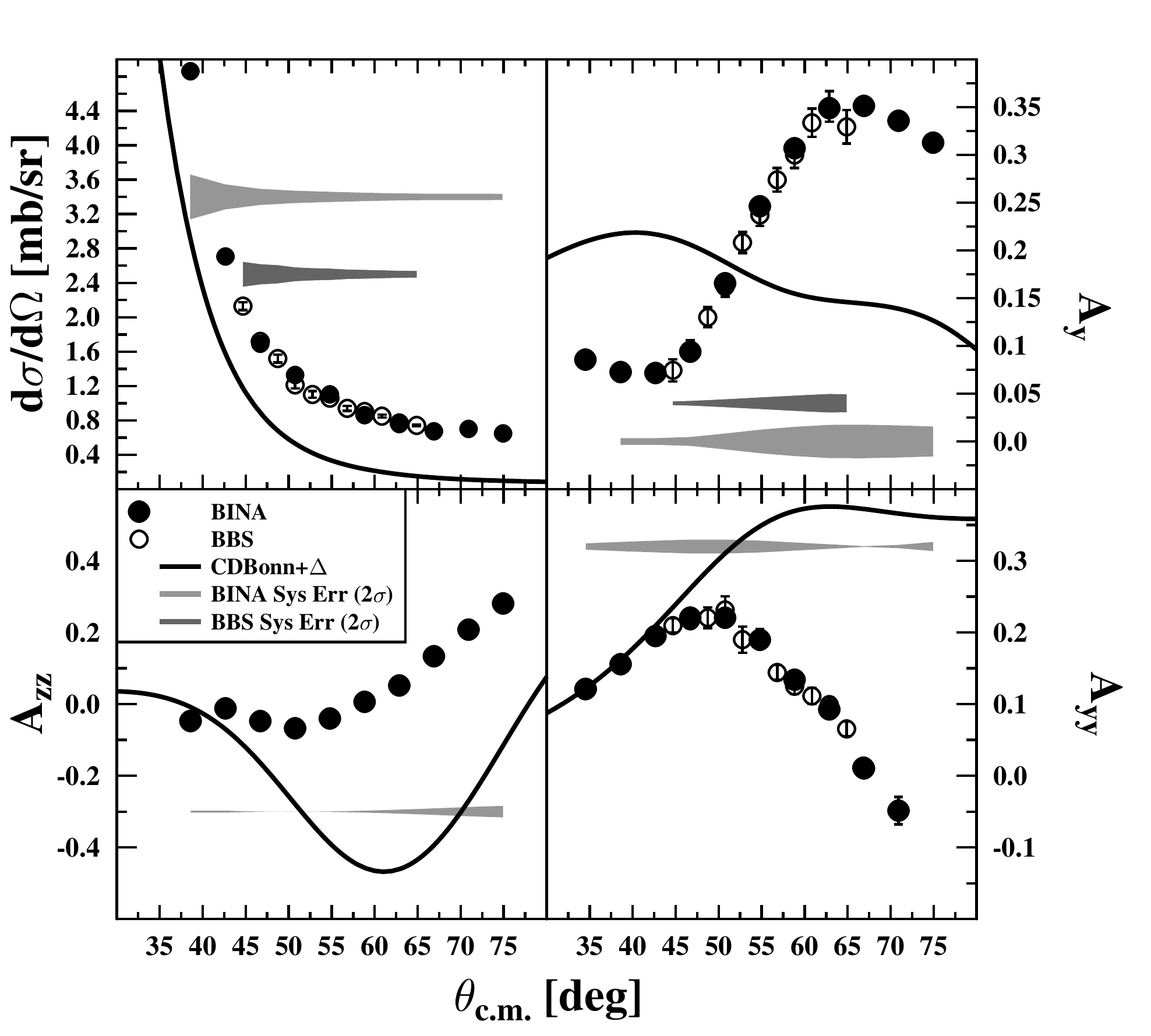}}
\vspace*{-0.2cm}
\caption{Differential cross section and analyzing powers of the elastic channel of the reaction $^{2}{\rm H}(\vec d,d)d$ are shown with statistical errors for each point. The total systematic uncertainty related to BINA (BBS) results is shown with a light (dark) gray band for each panel. The results of BINA data are shown as filled circles and those for the BBS data are presented by open circles~{\cite{BBSref}}. The solid lines are the result of a calculation based on the lowest-order terms in the Born series expansion of the Alt-Grassberger-Sandhas equation for a four-nucleons interaction using CD-Bonn+$\Delta$ potential~{\cite{Micher,Delt2,Delt3}}.}
\label{Crse}
\end{figure*}

\indent Figure~\ref{Crse} shows the measured differential cross sections and analyzing powers for the elastic deuteron-deuteron scattering, $^{2}{\rm H}(\vec d,d)d$. The results of BINA data are presented as filled circles and the results of data taken by BBS setup are shown as open circles~\cite{BBSref}. The light (dark) gray band in each panel shows the systematic uncertainty of the BINA (BBS) data, and the error bars represent the statistical errors which are smaller than the symbol size for most of the data points. As discussed before, the results of the $A_{yy}$ measurement obtained from BBS were used to normalize the offset of the current readout and hence, the corresponding systematic error is the same for both setups. Therefore, just one gray band is shown in Fig.~\ref{Crse}. The solid lines are the results of a rough
approximation based on the lowest-order terms in the Born series expansion of the Alt-Grassberger-Sandhas equation for a four-nucleons interaction
using CD-Bonn+$\Delta$ potential~{\cite{Micher,Delt2,Delt3}}. \\
\indent The comparison between the results of the two experiments, namely data taken from BINA and BBS setups, indicates that both data sets are in very good agreement within the uncertainties. But, comparing the experimental data with the theoretical approximation shows contradictions specially in the results of the analyzing powers. Aside from the normalization in the results of the differential cross section, the theoretical prediction follows at least the shape of the experimental data. In the case of analyzing powers, the comparison shows contradictory results indicating defects in the spin parts of theoretical calculations of the scattering amplitude. As already mentioned, the comparison between the results of exact calculations and those coming from Born approximation in Fig.~\ref{DP}, indicates that this approximation is not very suitable for the ${\rm H}(\vec d,dp)$ reaction in this energy range, and therefore, we expect to observe discrepancies between Born approximation and the experimental data in the $^{2}{\rm H}(\vec d,d)d$ reaction in Fig.~\ref{Crse}. In fact, Born approximation may provide a reasonable estimation for observables at higher energies and small angles, but, it is not reliable in the considered energy and angle regime in this paper. It indicates that exact theoretical calculations of four-body systems are a necessity to do a reasonable comparison with the experimental data.
\section{Summary}
\indent In summary, we have analyzed the elastic channel of deuteron-deuteron scattering, $^{2}{\rm H}(\vec d,d)d$, at 65~MeV/nucleon. Two experiments were performed with two independent setups, namely BINA and BBS, which were located at KVI in Groningen, the Netherlands. Cross sections and analyzing powers were obtained for a large angular range of the
phase space.  An excellent agreement is found between the measured differential cross sections and analyzing powers of both experiments for the angular range at which they overlap. The experimental results are also compared with a theoretical approximation based on lowest-order terms in the Born series expansion using CD-Bonn+$\Delta$ potential. The very poor agreement between the experimental data and theoretical approximations shows the necessity of {\it ab-initio} calculations in the four-body systems at intermediate energies.\\
\section{ACKNOWLEDGMENT}
The authors acknowledge the work by the cyclotron and ion-source groups at KVI for delivering a high-quality beam used
in these measurements. The present work has been performed with financial support from the ``Nederlandse Organisatie voor Wetenschappelijk Onderzoek'' (NWO). This work was partly supported by Iran National Science Foundation (INSF) as a research project under No 98028747. 
%

\end{document}